\documentclass[smus]{snow2e}
\usepackage{epsf}
\begin{document}
%
% Input some definitions
%
%
%----------  PHYSICS COMMANDS
%
\def \Et {{\rm E}_{\rm T}}
\def \Pt {{\rm P}_{\rm T}}
\def \enu {\epsilon_{\nu}}
\def \stw {$\sin^{2}\theta_{W}$}
\newcommand{\MET}{\mbox{$\protect \raisebox{.3ex}{$\not$}\et$}}
\def\deg{^\circ}
\def\qbar{{\bar q}}
\def\nubar{{\bar \nu}}
\def\W{{\em W\/ }}
\def\Z0{${\em Z^0\/}$}
\def\lum{{\cal L}}
\def\epem{{\rm e^{+}e^{-}}}
\def\tptm{{\tau^{+}\tau^{-}}}
\def\roots{${\sqrt s}\:$}
\def\r#1 {$^{#1}$}
\def\sigW {$\sigma\cdot$B(\W$\rightarrow~$e $\nu$) }
\def\sigZ {$\sigma\cdot$B(\Z0$\rightarrow~\epem$) }
\hyphenation{brem-sstrah-lung proc-ess}
\newcommand{\et}{{\rm E}_{\scriptscriptstyle\rm T}}
\newcommand{\etcone}{{\rm E}_{\scriptscriptstyle\rm T}^{cone}}
\newcommand{\abseta}{\mid \eta^{det} \mid \leq}
\newcommand{\abz}{\mid z \mid \leq}
\newcommand{\fb}{f_{b}}
\newcommand{\ks}{K_{s}^{0}}
\newcommand{\pich}{\pi^{\pm}}
\newcommand{\piz}{ \pi^{0} }
\newcommand{\bigz}{{\cal Z}}
\newcommand{\emf}{f_{em}}
\newcommand{\deltar}{\sqrt{\Delta \eta ^{2}+ \Delta \phi ^{2}}}
\newcommand{\etprime}{{\rm E}_{\scriptscriptstyle\rm T'}}
\newcommand{\ptran}{{\rm P}_{\scriptscriptstyle\rm T}}
\newcommand{\met}{\mbox{$\protect \raisebox{.3ex}{$\not$}\et$}}
\newcommand{\wenu}{W \rightarrow e \nu}
\newcommand{\wmunu}{W \rightarrow \mu \nu}
\newcommand{\wlep}{W \rightarrow \rm{lepton}\, \nu}
\newcommand{\zv}{{\rm z}_{vertex}}
\newcommand{\ppbar}{p\bar{p}}
\newcommand{\qqbar}{q\bar{q}}
\newcommand{\ttbar}{t\bar{t}}
\newcommand{\bbbar}{b\bar{b}}
\newcommand{\ccbar}{c\bar{c}}
\newcommand{\ppbb} { \ppbar \rightarrow  \bbbar }
\newcommand{\zee}{Z \rightarrow e^{+}e^{-} }
\newcommand{\bele}{b \rightarrow c e \nu_{e} }         
\newcommand{\blnu}{b \rightarrow c l \nu_{l} }         
\newcommand{\mtran}{{\rm M}_{\scriptscriptstyle\rm T}}
\newcommand{\acceff}{\rm{A} \times \epsilon}
% dilepton symbols
\def \mc {\multicolumn}
\def \pb    {pb$^{-1} $}
\def \DeltaPhi {$\Delta \phi_{\ell\,\ell \ }$} 
\def \mtop {$M_{top} $}
\def \ztau   {$Z\rightarrow\tau\tau \:$}
\def \DeltaPhil {$\Delta \phi{(\MET,\ell) \ }$} 
\def \DeltaPhij {$\Delta \phi{(\MET,j) \ }$} 
\def \TTbar {$t\overline{t} \; \;$}
\def \dpemu {\Delta \phi_{e\mu}}
\def \Mt {M_{top}}
\def \mtenu  {M_{T}^{e\nu}}
\def \lum {\cal L}
\def \intlum {\int {\cal L} dt}
\def \Zee {Z^{0} \rightarrow e^{+}e^{-}}
\def \Zmumu {Z^{0} \rightarrow \mu^{+}\mu^{-}}
\def \emu {e \mu}  
\def \temux {\ttbar \rightarrow \emu + X}
\def \Ete {E_T^{e}}
\def \Ptmin {P_T^{min}}
\def \Ptmu {P_T^{\mu}}
\def \Etmiss {{\not}{E_T}}
% end of dilepton
%---------- UNITS, SYMBOLS
%
\newcommand{\imb}{ \mu {\rm b}^{-1} }
\newcommand{\inb}{ {\rm nb}^{-1} }
\newcommand{\ipb}{ {\rm pb}^{-1} }
\newcommand{\degs}{\mbox{$^{\circ}$}}
\newcommand{\gsim}{\mbox{\small$\stackrel{>}{\sim}$\normalsize}}
\newcommand{\lsim}{\mbox{\small$\stackrel{<}{\sim}$\normalsize}}
%
%---------- TYPE SETTING
%
\newcommand{\etal}{{\em et al.}}
\newcommand{\tableskip}{\vskip 5pt plus3pt minus1pt \relax}
\newcommand{\tindent}{\hskip 17pt}
\newcommand{\hfull}{\hspace*{\fill}}
\newcommand{\tline}{\protect\linebreak[4]\hfull}
\newcommand{\linespace}[1]{\protect\renewcommand{\baselinestretch}{#1}
  \footnotesize\normalsize}
%
%---------- Journal names
%
\newcommand{\prl}[1]{Phys. Rev. Lett {\bf #1}}
\newcommand{\prev}[1]{Phys. Rev. {\bf #1}}
\newcommand{\prd}[1]{Phys. Rev. D {\bf #1}}
\newcommand{\zs}[1]{Z. Phys. {\bf #1}}
\newcommand{\ncim}[1]{Nuovo Cim. {\bf #1}}
\newcommand{\plet}[1]{Phys. Lett. {\bf #1}}
\newcommand{\prep}[1]{Phys. Rep. {\bf #1}}
\newcommand{\rmp}[1]{Rev. Mod. Phys. {\bf #1}}
\newcommand{\nphy}[1]{Nucl. Phys. {\bf #1}}
\newcommand{\nim}[1]{Nucl. Instrumen. Meth. {\bf #1}}
%

%------------- Figure commands and macros
%
%
%  Called the same way epsffile is called.  Difference is it will center
%  the graphic in the page useing the center environment.
%
\def\gepsfcentered#1{
  \def\testit{#1}
  \def\lbracket{[}
  \ifx\testit\lbracket
    \let\dofilecmd=\gepsfwithopt
  \else
    \let\dofilecmd=\gepsfnoopt
  \fi
  \dofilecmd}

\def\gepsfnoopt#1{
  \begin{center}
  \leavevmode
  \epsffile{#1}
  \end{center}}

\def\gepsfwithopt#1 #2 #3 #4]#5{
  \begin{center}
  \leavevmode
  \gepsfmaxx=0.94\textwidth
  \epsffile[#1 #2 #3 #4]{#5}
  \end{center}}

%
%  Auto sizing for epsf figures that are larger than the text width.
%
\newdimen\gepsfmaxx
\gepsfmaxx=0.94\textwidth
\def\epsfsize#1#2{
  \ifnum \epsfxsize=0
    \ifnum \epsfysize=0
      \ifnum #1 > \gepsfmaxx
        \gepsfmaxx
        %\message{Did scaling.}
      \else
        #1
        %\messaeg{Used nat scaling}
      \fi
    \else
      \epsfxsize
      %\message{Using what ever.}
    \fi
  \else
    \epsfxsize
    %\message{Again, using whatever.}
  \fi
  %\message{Hi epsfxsize is \the\epsfxsize ...}
  %\message{epsfysize is \the\epsfysize ...}
  %\message{Hi first arg is \the#1 ...}
  %\message{Second arg is \the#2 ...}
}

\title{Top Quark Physics Results from CDF and D0}

\author{David Gerdes\\ {\it Department of Physics and Astronomy, The Johns 
                         Hopkins University}\\ {\it 3400 North Charles Street,
                         Baltimore, MD 21218 USA}\\ 
                         {\it E-mail: gerdes@jhu.edu} }
\maketitle

\thispagestyle{plain}\pagestyle{plain}

\begin{abstract} 
I summarize recent top quark physics results from the Fermilab Tevatron
experiments. Since the observation of the top quark by CDF and D0 in
1995, the experimental focus has shifted to a detailed study of the
top quark's properties. This article describes recent measurements of 
the top quark production cross section, mass, kinematic properties, 
branching ratios, $V_{tb}$,  and the $W$ polarization in top decays.
\end{abstract}

\section{Introduction}
The existence of the top quark, which is required in the Standard Model
as the weak isospin partner of the bottom quark, was firmly established
in 1995 by the CDF\cite{cdf_obs} and D0\cite{d0_obs} experiments at the
Fermilab Tevatron, confirming earlier evidence presented by CDF\cite{
top_evidence_prd,top_kin_prd}. Each experiment reported a roughly 5$\sigma$
excess of $t\bar{t}$ candidate events over background, together with a 
peak in the mass distrbution for fully reconstructed events. The datasets
used in these analyses were about 60\% of the eventual Run I totals.
With the top quark well in hand and over 100~pb$^{-1}$ of data collected 
per experiment, the emphasis has now shifted to a more precise study of the 
top quark's properties.

In $p\bar{p}$ collisions at $\sqrt{s}=1.8$~TeV, the dominant top quark 
production mechanism is pair production through $q\bar{q}$ annihilation. In
the Standard Model, each top quark decays immediately to a $W$ boson and 
a $b$ quark. The observed
event topology is then determined by the decay mode of the two $W$'s. Events
are classified by the number of $W$'s that decay leptonically. About 5\%
of the time each $W$ decays to $e\nu$ or $\mu\nu$ (the ``dilepton channel''), 
yielding a final state with two isolated, high-$P_T$ charged leptons, 
substantial missing transverse energy ($\Etmiss$) from the undetected 
energetic neutrinos, and two $b$ quark jets. This final state is extremely
clean but suffers from a low rate. The ``lepton + jets'' final state
occurs in the 30\% of $t\bar{t}$ decays where when one $W$ decays to leptons 
and the other decays into quarks. These events contain a single high-$P_T$ 
lepton, large $\Etmiss$, and (nominally) four jets, two of which are from
$b$'s. Backgrounds in this channel can be reduced to an acceptable level
through $b$-tagging and/or kinematic cuts, and the large branching ratio
to this final state makes it the preferred channel for studying the top
quark at the Tevatron.
The ``all-hadronic'' final state occurs when both $W$'s decay to
$q\bar{q}^{\prime}$, which happens 44\% of the time. This final state contains
no leptons, low $\Etmiss$, and six jets, including two $b$ jets. Although
the QCD backgrounds in this channel are formidable, extraction of the signal
is possible through a combination of $b$-tagging and kinematic cuts. Finally,
approximately 21\% of $t\bar{t}$ decays are to final states containing 
$\tau$'s. Backgrounds to hadronic $\tau$ decays are large, and while signals
have been identified I will not discuss these analyses here.

This paper is organized as follows. Section~\ref{sec-xsection} discusses
the measurement of the $t\bar{t}$ production cross section. The measurement
of the top quark mass is described in Section~\ref{sec-mass}. Kinematic
properties of $t\bar{t}$ production are described in Section~\ref{sec-kine}.
The measurement of the top quark branching ratio to $Wb$ and the CKM matrix
element $V_{tb}$ is described in Section~\ref{sec-Vtb}. Section~\ref{sec-rare}
discusses searches for rare or forbidden decays of the top. 
Section~\ref{sec-Wpol} discusses a measurement of the $W$ polarization in
top decays. Section~\ref{sec-concl} concludes.

\section{Production Cross Section}
\label{sec-xsection}

The measurement of the top quark production cross section $\sigma_{t\bar{t}}$ 
is of interest for a number of
reasons. First, it checks QCD calculations of top production, which have
been performed by several groups\cite{xsec-berger,xsec-catani,xsec-laenen}.
Second, it provides an important benchmark for estimating
top yields in future high-statistics experiments at the Tevatron and LHC.
Finally, a value of the cross section significantly different from the QCD
prediction
could indicate nonstandard production or decay mechanisms, for example
production through the decay of an intermediate high-mass state or decays
to final states other than $Wb$.

\subsection{CDF Measurements of $\sigma_{\ttbar}$}

The CDF collaboration has measured the $\ttbar$ production cross section
in the dilepton and lepton + jets modes, and in addition has recently
performed a measurement in the all-hadronic channel. The dilepton and
lepton + jets analyses begin with a common inclusive lepton sample,
which requires an isolated electron or muon with $P_T > 20$~GeV and
$|\eta|<1$. The integrated luminosity of this sample is 110~pb$^{-1}$.

For the dilepton analysis, a second lepton is required with $P_T>20$~GeV.
The second lepton must have an opposite electric charge to the primary lepton
and may satisfy a looser set of identification cuts. In addition, two
jets with $E_T>10$~GeV are required, and the $\Etmiss$ must be greater
than 25~GeV. For the case $25 < \Etmiss < 50$~GeV, the $\Etmiss$ vector must
be separated from the nearest lepton or jet by at least 20 degrees. This
cut rejects backgrounds from $Z\rightarrow \tau\tau$ decays followed
by $\tau\rightarrow (e~{\rm or}~\mu)$ (where the
$\Etmiss$ tends to lie along the lepton direction) and from events containing
poorly measured jets (where the $\Etmiss$ tends to lie along a jet axis).
Events where the dilepton invariant mass lies between 75 and 105 GeV are
removed from the $ee$ and $\mu\mu$ channels as $Z$ candidates. In addition,
events containing a photon with $E_T > 10$ GeV are removed if the $ll\gamma$
invariant mass falls within the $Z$ mass window. This ``radiative $Z$'' cut
removes one event from the $\mu\mu$ channel and has a negligible effect
on the $\ttbar$ acceptance and backgrounds. Nine dilepton candidates
are observed: one $ee$, one $\mu\mu$, and seven $e\mu$ events. Including
a simulation of the trigger acceptance, the expected division of dilepton
signal events is 58\% $e\mu$, 27\% $\mu\mu$, and 15\% $ee$, consistent with
the data. It is also interesting to note that four of the nine
events are $b$-tagged, including two double-tagged events. Although no 
explicit $b$-tag requirement is made in the dilepton analysis, the fact that
a large fraction of the events are tagged is powerful additional evidence
of $\ttbar$ production. 

Backgrounds in the dilepton channel arise from Drell-Yan production of
lepton pairs, diboson production, $Z\rightarrow\tau\tau$, $b\bar{b}$, and
fakes. These backgrounds are estimated through a combination of data and
Monte Carlo. The total background in the $ee + \mu\mu$ channels is 
$1.21\pm 0.36$ events, and is $0.76\pm 0.21$ events in the $e\mu$ channel.
Event yields, backgrounds, and estimated $\ttbar$ contributions are 
summarized in Table~\ref{tab-cdf_dil}. 

When these numbers are combined with the $\ttbar$ acceptance in the
dilepton mode of $0.78\pm0.08$\% (including branching ratios), and using
CDF's measured top mass of 175~GeV (described below), the resulting
cross section is $\sigma_{\ttbar} = 8.3^{+4.3}_{-3.3}$~pb.

\begin{table}[h]
\begin{center}
\caption{Summary of event yields and backgrounds in the CDF dilepton
analysis. Expected $\ttbar$ contributions are also shown.}
\label{tab-cdf_dil}
\begin{tabular}{ccc}
\hline
\hline
Background  & $ee, \mu\mu$ & $e\mu$  \\
\hline
Drell-Yan   &   $0.60\pm0.30$   & --- \\
$WW$        &   $0.16\pm0.07$   & $0.20\pm0.09$ \\
fakes       &   $0.21\pm0.17$   & $0.16\pm0.16$ \\
$b\bar{b}$  &   $0.03\pm0.02$   & $0.02\pm0.02$ \\
$Z\rightarrow\tau\tau$ & $0.21\pm0.08$ & $0.38\pm0.11$ \\ \hline
Total bkgd. &   $1.21\pm0.36$   & $0.76\pm0.21$ \\ \hline
Expected $\ttbar$, &  2.6, 1.6, 1.0 & 3.9, 2.4, 1.5 \\
$M_{top}=160,175,190$ &             &                \\ \hline
Data (110 pb$^{-1}$) & 2           &     7       \\ 
\hline
\end{tabular}
\end{center}
\end{table}

The lepton + jets cross section analysis begins with the common inclusive 
lepton sample described above. An inclusive $W$ sample is selected from
this sample by requiring $\Etmiss > 20$~GeV. Jets are clustered in a cone
of $\Delta R \equiv \sqrt{(\Delta\eta)^2 + (\Delta\phi)^2} = 0.4$, and at least
three jets with $E_T > 15$~GeV and $|\eta|<2$ are required in the $\ttbar$
signal region. (These jet energies are not corrected for detector effects,
out-of-cone energy, the underlying event, etc. Such corrections
are applied later, in the mass analysis. The average correction factor is
about 1.4.) $Z$ candidates are removed as before, and the lepton is
required to pass an appropriate trigger. Finally, the event is required
not to have been accepted by the dilepton analysis above. The dilepton
and lepton + jets samples are therefore nonoverlapping by construction.
There are 324 $W + \ge 3$-jet events in this sample. 

Signal to background in this sample is approximately 1:4. CDF employs
two $b$-tagging techniques to reduce background. 
The first technique identifies
$b$ jets by searching for a lepton from the decay $b\rightarrow l\nu X$ or
$b\rightarrow c\rightarrow l\nu X$. Since this lepton typically has
a lower momentum than the the lepton from the primary $W$ decay, 
this technique is known as the ``soft lepton tag'' or SLT. In addition
to tagging soft muons, as in the D0 analysis, CDF also identifies soft
electrons. The second, more powerful,
technique exploits the finite lifetime of the $b$ quark by searching for
a secondary decay vertex. Identification of these vertices 
is possible because of the excellent impact parameter resolution 
of CDF's silicon microstrip vertex detector, the SVX\cite{cdf_svx,cdf_svxp}.
This technique is known as the ``SVX tag.''

The SLT algorithm identifies electrons and muons from semileptonic $b$
decays by matching central tracks with electromagnetic energy clusters
or track segments in the muon chambers. To maintain acceptance for leptons
coming from both direct and sequential decays, the $P_T$ threshold
is kept low (2~GeV). The fiducial region for SLT-tagged leptons is
$|\eta|<1$. The efficiency for SLT-tagging a $\ttbar$ event is $20\pm2$\%,
and the typical fake rate per jet is about 2\%. The details of the SLT
algorithm are discussed in Ref.~\cite{top_evidence_prd}.

The SVX algorithm begins by searching for displaced vertices containing
three or more tracks which satisfy a ``loose'' set of track quality 
requirements. Loose track requirements are possible because the probability
for three tracks to accidentally intersect at the same displaced space
point is extremely low. If no such vertices are found, two-track vertices
that satisfy more stringent quality cuts are accepted. A jet is defined to 
be tagged if it contains a secondary vertex whose transverse displacement
(from the primary vertex) divided by its uncertainty is greater than three.
The efficiency for SVX-tagging a $\ttbar$ event is $41\pm 4$\%, nearly
twice the efficiency of the SLT algorithm, while the fake rate is
only $\simeq 0.5$\% per jet. The single largest source of inefficiency
comes from the fact that the SVX covers only about 65\% of the Tevatron's
luminous region. SVX-tagging is CDF's primary $b$-tagging technique.

Table~\ref{tab:cdf_ljets_btag} summarizes the results of tagging in 
the lepton + jets sample. The signal region is $W +\ge3$ jets, where there
are 42 SVX tags in 34 events and 44 SLT tags in 40 events, on backgrounds
of $9.5\pm1.5$ and $23.9\pm2.8$ events respectively. SVX backgrounds are
dominated by real heavy flavor production ($Wb\bar{b}$, $Wc\bar{c}$, $Wc$),
while SLT backgrounds are dominated by fakes. Monte Carlo calculations are
used to determine the fraction of observed $W$+jets events that contain
a heavy quark, and then the observed tagging efficiency is used to derive
the expected number of tags from these sources. Fake rates are measured 
in inclusive jet data. Backgrounds are corrected iteratively for the assumed
$\ttbar$ content of the sample. 

When combined with the overall $\ttbar$ acceptance in the lepton + jets
mode, $\sigma_{\ttbar}$ is measured to be $6.4^{+2.2}_{-1.8}$~pb using SVX
tags, and $8.9^{+4.7}_{-3.8}$~pb using SLT tags.

\begin{table}[htbp]
\begin{center}
\caption{Summary of results from the CDF lepton + jets $b$-tag analysis. 
The expected $\ttbar$ contributions are calculated using CDF's measured
combined cross section.}
\begin{tabular}{lccc}
\hline
\hline
           &   $W$ + 1 jet   & $W$+2 jets & $W+\ge$3 jets \\ \hline
Before tagging & 10,716      & 1,663      & 324 \\ \hline
SVX tagged evts & 70       &  45        &  34 \\
SVX bkgd    & $70\pm11$& $32\pm4$   & $9.5\pm1.5$ \\
Expected $\ttbar$ & $0.94\pm0.4$ & $6.4\pm2.4$ & $29.8\pm8.9$
                                                             \\ \hline
SLT tagged evts & 245      &  82         & 40          \\
SLT bkgd   & $273\pm24$& $80\pm6.9$ & $23.9\pm2.8$ \\
Expected $\ttbar$ & $1.1\pm0.4$  & $4.7\pm1.6$ & $15.5\pm5.3$ \\
\hline
\end{tabular}
\label{tab:cdf_ljets_btag}
\end{center}
\end{table}

CDF has also performed a measurement of $\sigma_{\ttbar}$ in the
all-hadronic channel, which nominally contains six jets, no leptons,
and low $\Etmiss$.
 Unlike in the case of lepton + jets, $b$-tagging
alone is not sufficient to overcome the huge backgrounds from QCD multijet
production. A combination of kinematic cuts and SVX $b$-tagging is therefore
used. 

The initial dataset is a sample of about 230,000 events containing
at least four jets with $E_T>15$~GeV and $|\eta|<2$. Signal to background
in this sample is a forbidding 1:1000, so a set of kinematic cuts is
applied. The jet multiplicity is required to be $5\le N_{jets}\le 8$,
and the jets are required to be separated by $\Delta R\ge 0.5$. Additionally,
the summed transverse energy of the jets is required to be greater than
300 GeV and to be ``centrally'' deposited: $\sum E_T(jets)/\sqrt{\hat{s}} >
0.75$, where $\sqrt{\hat{s}}$ is the invariant mass of the multijet system.
Finally, the $N_{jet}-2$ subleading jets are required to pass an aplanarity
cut. The resulting sample of 1630 events has a signal to background of about
1:15. After the requirement of an SVX tag, 192 events remain.

The tagging background is determined by appying the SVX tagging probabilities
to the jets in the 1630 events selected by the analysis prior to tagging.
The probabilities are measured from multijet events and are parametrized
as a function of jet $E_T$, $\eta$, and SVX track multiplicity. The
probability represents the fraction of jets which are tagged in the absence
of a $\ttbar$ component, and includes real heavy flavor as well as mistags.
Applying the tagging probabilities to the jets in the 1630 events remaining
after kinematic cuts, a predicted background of $137\pm 11$ events is obtained,
compared to the 192 tagged events observed.

The efficiency to SVX-tag a $\ttbar$ event in the all-hadronic mode
is $47\pm5$\%. This value is slightly larger than the lepton + jets case
due to the presence of additional charm tags from $W\rightarrow c\bar{s}$.
Combining this value with the acceptance for the all-hadronic mode, including
the efficiency of the multijet trigger and the various kinematic cuts,
CDF obtains a $\ttbar$ cross section in this channel of 
$10.7^{+7.6}_{-4.0}$~pb.

\begin{figure}[htbp]
\begin{center}
\leavevmode
\epsfysize=3.5in
\epsffile{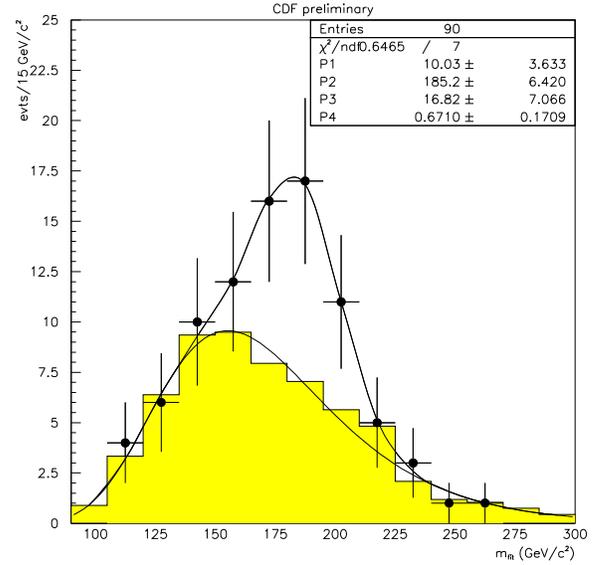}
\caption{Reconstructed top mass obtained from a constrained fit to SVX-tagged 
events in the CDF all-hadronic analysis.}
\label{fig:all_had_mass}
\end{center}
\end{figure}

The large background in the all-hadronic channel makes it desirable to
have some independent cross check that the observed excess of events is
really due to $\ttbar$ production. The events in this sample with exactly
six jets can be matched to partons from the process $\ttbar\rightarrow
WbW\bar{b}\rightarrow jjbjj\bar{b}$, and can be fully reconstructed.
A plot of the reconstructed top mass for these events is shown in 
Fig.~\ref{fig:all_had_mass}. The events clearly display a peak at the
value of the top mass measured in other channels. This analysis impressively
illustrates the power of SVX-tagging to extract signals from
very difficult environments.

\begin{figure}[htbp]
\begin{center}
\leavevmode
\epsfysize=3.5in
\epsffile{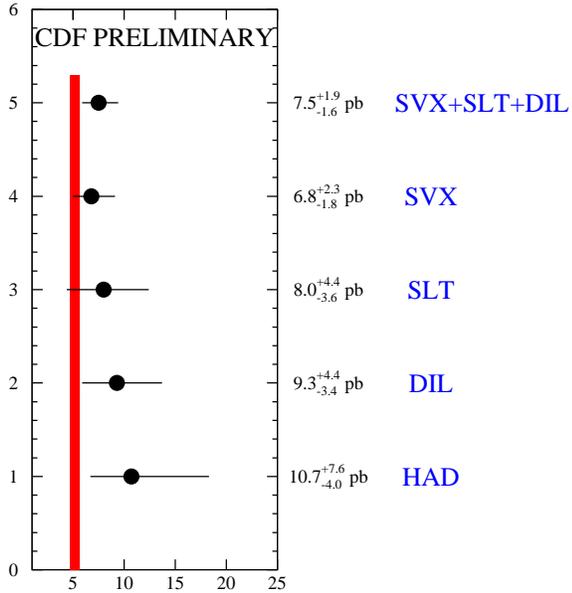}

\caption{CDF values of $\sigma_{\ttbar}$ for individual channels and for the 
combined measurement. The band represents the central value and spread of
the theoretical value from three recent calculations for $M_{top}=175$~GeV.}
\label{fig:xsec-lepstyle}
\end{center}
\end{figure}

The combined $\ttbar$ cross section is obtained using the number of events,
backgrounds, and acceptances for each of the channels. The calculation is
done using the likelihood technique described in Ref.~\cite{top_evidence_prd}.
Acceptances are calculated using $M_{top}=175$~GeV. The likelihood method
takes account of correlated uncertainties such as the luminosity uncertainty,
acceptance uncertainty from initial state radiation, etc. The combined
$\ttbar$ production cross section for $M_{top}=175$~GeV is 
\begin{equation}
       \sigma_{\ttbar} = 7.7^{+1.9}_{-1.6}~{\rm pb}~{\rm (CDF~Prelim.)}
\end{equation}
where the quoted uncertainty includes both statistical and systematic
effects. Fig.~\ref{fig:xsec-lepstyle} shows the individual and combined
CDF measurements together with the theretical central value and spread.
All measurements are in good agreement with theory, though all fall on
the high side of the prediction. It is perhaps noteworthy that the single
best measurement, from SVX-tagging in the lepton+jets mode, is the one
closest to theory.

\subsection{D0 Measurements of $\sigma_{\ttbar}$}
The D0 collaboration has measured $\sigma_{t\bar{t}}$ in both the dilepton 
($ee$, $e\mu$, and $\mu\mu$) and lepton + jets channels. The dilepton analysis
is a straightforward counting experiment. Two high-$P_T$ leptons are
required, as well as two jets. Cosmic ray and $Z$ candidates are removed.
In the $ee$ and $e\mu$ channels, a cut is
also placed on the missing transverse energy. Finally, a cut on $H_T$, the
transverse energy of the jets plus the leading electron (or the jets only,
in the case of dimuon events) is applied to reduce backgrounds from $W$ pairs,
Drell-Yan, etc. 
%The details of the cuts are shown in Table~\ref{tab-D0dil},
%together with the event yields and estimated backgrounds. 
%Table~\ref{tab-D0dil-ttbar} shows the predicted contribution in each channel
%from $\ttbar$, using the theoretical calculation of $\sigma_{t\bar{t}}$ from
%Ref.~\cite{xsec-laenen}. 
The largest acceptance is in the $e\mu$ channel,
which also has the lowest backgrounds. Three candidate events are observed
in this channel on a background of $0.36\pm 0.09$ events. For $M_{top}$ = 180
GeV, $1.69\pm 0.27$ signal events are expected in this channel.
One event is
observed in each of the $ee$ and $\mu\mu$ channels on backgrounds of
$0.66\pm 0.17$ and $0.55\pm 0.28$ events respectively. For $M_{top}$ = 180
GeV, one expects $0.92\pm0.11$ and $0.53\pm 0.11$ $\ttbar$ in these two
channels.

The D0 measurement of $\sigma_{\ttbar}$ in the lepton + jets channel
makes use of two different approaches to reducing the background from
$W$+jets and other sources: topological/kinematic cuts, and $b$-tagging.
The first approach exploits the fact that the large top quark mass 
gives rise to kinematically distinctive events:
the jets tend be more energetic and more central than jets in
typical background events, and the events as a whole are more spherical. 
Top-enriched samples can therefore be selected
with a set of topological and kinematic cuts. (For some earlier work on
this subject, see Refs.~\cite{top_kin_prd} and \cite{cdf_Ht}.) In particular,
the total hadronic activity in the event, $H_T \equiv \sum E_T(jets)$, 
can be combined with the aplanarity of the $W$ + jets 
system to reduce
backgrounds substantially. Cuts on both of these variables were used in
the original D0 top discovery analysis\cite{d0_obs}, and these cuts have 
now been reoptimized on Monte Carlo samples for use in the cross section 
measurement. A third kinematic variable with discriminating power, the
total leptonic transverse energy ($E_T^L\equiv E_T^{lep} + \Etmiss$) is
also used. Events are required to have four jets with $E_T>15$~GeV and
$|\eta| < 2.$
In 105.9~pb$^{-1}$ of $e,\mu$ + jet data, a total of 21 candidate
events are observed, on a background of 9.23$\pm$ 2.83 events that is dominated
by QCD production of $W$ + jets. For comparison, 19$\pm$ 3 (13 $\pm$ 2) 
events are expected
for $M_{top} = 160$~(180) GeV, again using the theoretical cross section from
Ref.~\cite{xsec-laenen}.

\begin{figure}[ht]
\begin{center}
\epsfysize=3.5in
\leavevmode\epsffile{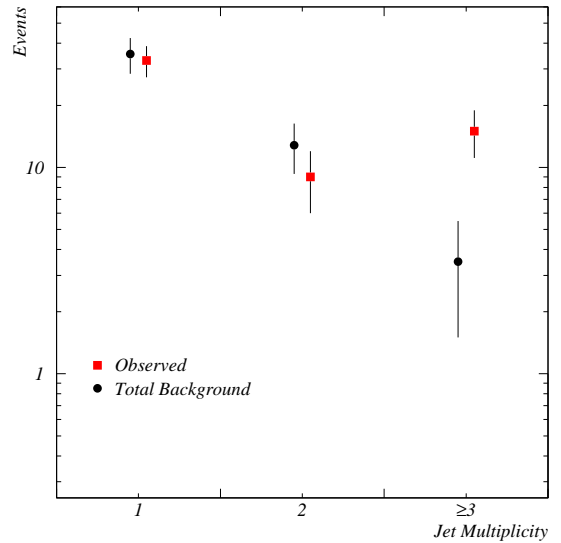}
\end{center}
\caption{Number of observed ($e,\mu$) + jets events with a soft muon tag
compared to background predictions, as a function of jet multiplicity.
Note the excess in the $\ttbar$ signal region with $W + \ge 3$ jets.}
\label{fig-d0_njets_mutag}
\end{figure}

A second D0 approach to the lepton + jets cross section measurement makes
use of $b$-tagging via soft muon tags. Soft muons are expected to be produced
in $\ttbar$ events through the decays $b\rightarrow\mu X$ and 
$b\rightarrow c\rightarrow\mu X$. Each $\ttbar$ event contains two $b$'s,
and ``tagging muons'' from their semileptonic decays are detectable in
about 20\% of $\ttbar$ events. Background events, by contrast, contain a low
fraction of $b$~quarks and thus produce soft muon tags at only the $\sim 2$\%
level. Events selected
for the lepton + jets + $\mu$-tag analysis are required to contain an $e$
or $\mu$ with $E_T$ ($P_T$ for muons) $>~20$ GeV, and to have $|\eta|<2.0~
(1.7)$ respectively. At least three jets are required with $E_T> 20$~GeV
and $|\eta|<2.$ The $\Etmiss$ is required to be at least 20~GeV (35 GeV
if the $\Etmiss$ vector is near the tagging muon in an $e$+jets event), and
in $\mu$ + jets events is required to satisfy certain topological cuts 
aimed at rejecting backgrounds from fake muons. Loose cuts on the
aplanarity and $H_T$ are also applied. Finally, the tagging muon is
required to have $P_T > 4$~GeV and to be near one of the jets, as would
be expected in semileptonic $b$ decay. In 95.7 pb$^{-1}$ of $e, \mu$ + 
jet data with a muon tag, 11 events are observed on a background ($W$ + jets,
fakes, and residual $Z$'s) of 2.58$\pm$0.57 events. Theory\cite{xsec-laenen}
predicts 9.0$\pm$2.2 and 5.2$\pm$1.2 events for $M_{top}$ = 160 and 180 GeV
respectively. Figure~\ref{fig-d0_njets_mutag} shows the clear excess of 
events in the signal region compared to the top-poor regions of one and 
two jets.

Table~\ref{tab-d0_xsec_summary} summarizes event yields and backgrounds 
in the D0 cross
section analysis. A total of 37 events is observed in the various
dilepton and lepton + jets channels on a total background of 13.4$\pm$3.0
events. The expected contribution from $\ttbar$ ($M_{top}$ = 180~GeV) is
21.2$\pm$ 3.8 events.

\begin{table}[h]
\begin{center}
\caption{Summary of event yields and backgrounds in the D0 cross section
analysis. Expected $\ttbar$ contributions are calculated for 
$M_{top}=180$~GeV.}
\label{tab-d0_xsec_summary}
\begin{tabular}{ccccc}
\hline
\hline
Channel  & $\int{\cal L}\, dt$ & Bkgd. & Expected $\ttbar$ & Data \\
\hline
$e\mu$  & 90.5          & $0.36\pm0.09$ & $1.69\pm0.27$ & 3 \\
$ee$    & 105.9         & $0.66\pm0.17$ & $0.92\pm0.11$ & 1 \\
$\mu\mu$& 86.7          & $0.55\pm0.28$ & $0.53\pm0.11$ & 1 \\ \hline
$e$+jets& 105.9         & $3.81\pm1.41$ & $6.46\pm1.38$ & 10 \\
$\mu$+jets& 95.7        & $5.42\pm2.05$ & $6.40\pm1.51$ & 11 \\
$e$+jets/$\mu$& 90.5    & $1.45\pm0.42$ & $2.43\pm0.42$ & 5 \\
$\mu$+jets/$\mu$& 95.7  & $1.13\pm0.23$ & $2.78\pm0.92$ & 6 \\ \hline
Total   &               & $13.4\pm3.0$  & $21.2\pm3.8$  & 37 \\

\hline
\end{tabular}
\end{center}
\end{table}

\begin{figure}[ht]
\begin{center}
\epsfysize=3.5in
\leavevmode\epsffile{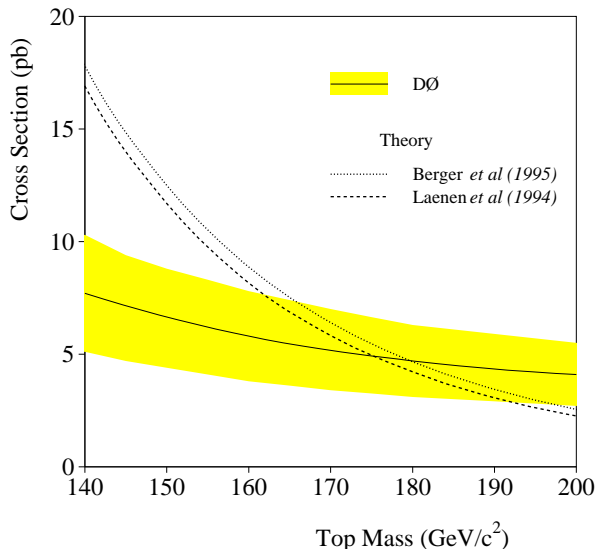}
\end{center}
\caption{D0 measurement of the $\ttbar$ production cross section as a
function of $M_{top}$.}
\label{fig-d0_xsec}
\end{figure}

When combined with a Monte Carlo calculation of the $\ttbar$ acceptance, these
numbers can be converted into a measurement of the cross section. 
Figure~\ref{fig-d0_xsec} shows the cross section derived from D0 data as
a function of $M_{top}$. For D0's measured top mass of 170 GeV, described
below, the measured $\ttbar$ cross section is 
\begin{equation}
      \sigma_{\ttbar} = 5.2\pm 1.8~{\rm pb}~{\rm (D0~Prelim.)}, 
\end{equation}
in good agreement with theory.

\section{Top Quark Mass Measurement}
\label{sec-mass}

The top quark mass is a fundamental parameter in the Standard Model.
It plays an important role in radiative corrections that relate electroweak
parameters, and when combined with other precision electroweak data can
be used to probe for new physics. In particular, the relationship between
$M_W$ and $M_{top}$ displays a well-known dependence on the mass
of the Higgs. A precise measurement of the top mass is therefore a high
priority of both experiments. 

The primary method for measuring the top mass at the Tevatron is a constrained
fit to lepton + 4-jet events arising from the process $\ttbar\rightarrow
WbW\bar{b}\rightarrow l\nu jjb\bar{b}$. 
In these events, the observed
particles and $\Etmiss$ can be mapped one-to-one to partons from the
$\ttbar$ decay. However, there are 12 possible
jet--parton assignments. The number of jet combinations is reduced to six 
if one $b$-tag is present, and to two if two $b$'s are tagged.
To select the best combination, both experiments use a likelihood
method that exploits the many constraints in the system. Each event is
fitted individually to the hypothesis that three of the jets come from 
one $t$ or $\bar{t}$ through its decay to $Wb$, and that the lepton,
$\Etmiss$, and the remaining jet come from the other $t$ or $\bar{t}$
decay. The fit is performed for each jet combination, with the requirement
that any tagged jets must be assigned as $b$ quarks in the fit. Each
combination has a two-fold ambiguity in the longitudinal momentum of the
neutrino. CDF chooses the solution with the best $\chi^2$, while D0
takes a weighted average of the three best solutions. In both cases,
solutions are required to satisfy a $\chi^2$ cut. The result is
a distribution of the best-fit top mass for each of the candidate events.
The final value for the top mass is extracted by fitting this distribution
to a set of Monte Carlo templates for $\ttbar$ and background. A likelihood
fit is again used to determine which set of $\ttbar$ templates best fits
the data. Because this measurement involves precision jet spectroscopy,
both experiments have developed sophisticated jet energy corrections,
described below, that relate measured jet energies to parton 
four-vectors. Uncertainties associated with these corrections are the
largest source of systematic error.

Measurements of the top mass in other channels (dilepton, all-hadronic\dots
have larger uncertainties, and give results consistent with the lepton + jets
measurements. These channels will not be discussed here. I now describe the 
CDF and D0 measurements in more detail.

\subsection{D0 Measurement of $M_{top}$}

\begin{figure}[htbp]
\begin{center}
\leavevmode
\epsfysize=3.5in
\epsffile{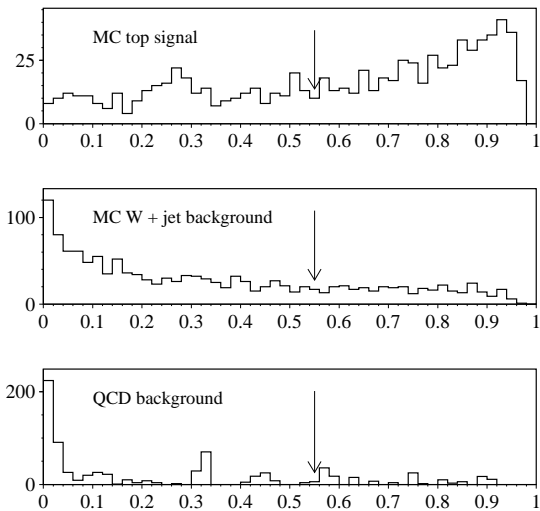}
\caption{Top likelihood distributions for $e$+jets signal and
background Monte Carlo events. 
The D0 top mass analysis uses events with top likelihood $> 0.55$.}
\label{fig:D0_top_like}
\end{center}
\end{figure}

\begin{figure}[htbp]
\begin{center}
\leavevmode
\epsfysize=3.2in
\epsffile{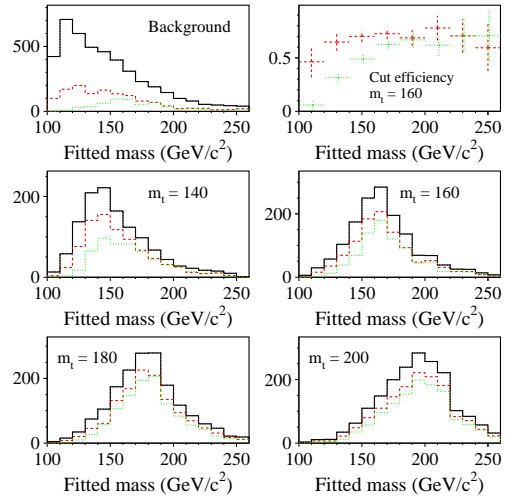}
\caption{Fitted mass distributions for background events and $\ttbar$ events 
of various masses in the D0 analysis. Histogram: parent sample. Dot-dash: 
after top likelihood cut. Dots: after cross-section cuts. Note the smaller 
bias introduced by the likelihood cut.}
\label{fig:D0_templates}
\end{center}
\end{figure}

The D0 top mass measurement begins with event selection cuts similar to
those used in the lepton + jets cross section analysis, with two important
differences. First, all events are required to have at least four jets with
$E_T > 15$~GeV and $|\eta|<2$. (Recall that in the cross section analysis,
soft-muon tagged events were allowed with only three jets.) Second and
more importantly, the cut on the total hadronic $E_T$ ($\equiv H_T$), which 
proved extremely useful for selecting a high-purity sample in the cross
section analysis, is replaced by a new ``top likelihood'' cut that combines
several kinematic variables. A straightforward $H_T$ cut would inject
significant bias into the analysis by pushing both background and signal
distributions toward higher values of $M_t$ and making background look like
signal. The top likelihood variable combines the $\Etmiss$, the aplanarity
of the $W$ + jets, the fraction of the $E_T$ of the $W$ + jets system that
is carried by the $W$, and the $E_T$-weighted rms $\eta$ of the $W$ and jets.
The distributions for each of these variables are determined from $\ttbar$
Monte Carlo events, and the probabilities are combined such that the bias
of the fitted mass distributions is minimized. The top likelihood distributions
for signal and background Monte Carlo events are shown in 
Fig.~\ref{fig:D0_top_like}. The advantages of this variable are demonstrated
in Fig.~\ref{fig:D0_templates}, which compares fitted mass distributions
for signal and background Monte Carlo events after the likelihood cut and
after the cross section ($H_T$) cuts. The top likelihood cut gives a
significantly smaller shift in the fitted distributions. This is particularly
true in the case of background events, where the cross section cuts ``sculpt''
the background distribution into a shape that looks rather top-like. 
The reduction of this source of bias is particularly important since
the D0 top mass sample is nearly 60\% background. A total
of 34 events pass the selection cuts, of which 30 have a good fit to the 
$\ttbar$ hypothesis.

For reconstructing the top mass, one desires to know the four-momenta of
the underlying partons as accurately as possible. In practice one observes 
jets, usually reconstructed with a fixed-cone algorithm, and several effects
can complicate the connection between these jets and their parent partons.
Calorimeter nonlinearies, added energy from multiple interactions and the
underlying event, uranium noise in the calorimeter, 
and energy that falls outside of the jet clustering cone all must be accounted
for. The D0 jet corrections are derived from an examination of events in
which a jet recoils against a highly electromagnetic object (a ``$\gamma$'').
The energy of the ``$\gamma$'' is well-measured in the electromagnetic 
calorimeter, whose energy scale is determined from $Z\rightarrow ee$ events.
It is then assumed that the component of the $\Etmiss$ along the jet axis 
($\Etmiss_{,\parallel}$) is
due entirely to mismeasurement of the jet energy, and a correction factor for
the recoil jet energy is obtained by requiring $\Etmiss_{,\parallel}$ to
vanish. The correction factors are derived as a function of jet $E_T$ and
$\eta$.

These jet corrections are ``generic'' and are used in many D0 analyses,
including the $\ttbar$ cross section analysis. Additional corrections are
applied for the top mass analysis. These corrections account for the fact
that light quark jets (from hadronic $W$ decays) and $b$ quark jets have
different fragmentation properties. Furthermore, $b$ jets tagged with the
soft muon tag must have the energy of the minimum-ionizing muon added back
in, and a correction must be applied for the neutrino. These flavor-dependent
corrections are determined from $\ttbar$ Monte Carlo events. The flavor 
assignment of the jets is established by the constrained fit. 

Backgrounds in the 30-event final sample come from the QCD production of
$W$ + multijets, and from fakes. These backgrounds are calculated for each 
channel before the top likelihood cut. The effects of the top likelihood cut 
and the fitter $\chi^2$ cut are determined from Monte Carlo. The result
is an estimated background of 17.4$\pm$2.2 events. The background is 
constrained to this value (within its Gaussian uncertainties) in the
overall fit to $\ttbar$ plus background templates that determines the
most likely top mass. 

\begin{figure}[htbp]
\begin{center}
\leavevmode
\epsfysize=3.2in
\epsffile{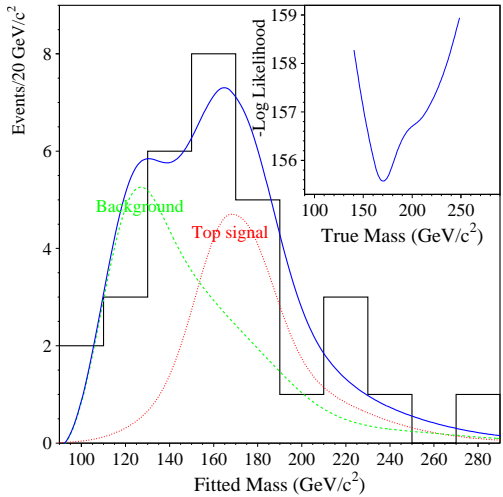}
\caption{Reconstructed top mass distribution from D0 data, together with results of the best fit.}
\label{fig:D0_mtfit}
\end{center}
\end{figure}

Figure~\ref{fig:D0_mtfit} shows the reconstructed mass distribution for the
30 events, together with the results of the fit. The result is
$M_{top} = 170\pm 15$(stat)~GeV. The statistical error is determined
by performing a large number of Monte Carlo ``pseudo-experiments'' with
$N=30$ events and $\bar{N}_{bkgd}=17.4$. The standard deviation of the 
mean in this ensemble of pseudo-experiments is taken to be the statistical 
error.

Systematic uncertainties come from the determination of the jet energy scale
from $Z\rightarrow ee$ events ($\pm$7 GeV), variations among Monte Carlo 
generators ({\sc Isajet} vs. the default {\sc Herwig}) and jet 
definitions ($\pm$6 GeV), uncertainties in the background
shape ($\pm$ 3 GeV), variations in the likelihood fitting method ($\pm$ 3
GeV), and Monte Carlo statistics ($\pm$ 1 GeV). The final result is therefore
\begin{equation}
      M_{top}= 170\pm 15 ({\rm stat})\pm 10 ({\rm syst})~{\rm GeV}~({\rm 
                                                          D0~prelim.}) 
\end{equation}

\subsection{CDF Measurement of $M_{top}$}

At the winter `96 conferences and at Snowmass, CDF reported a top mass 
value of $M_{top} = 175.6\pm 5.7({\rm stat})\pm 7.1({\rm syst})$~GeV.
This value was obtained using a technique very similar to that reported
in Refs.~\cite{cdf_obs} and~\cite{top_evidence_prd}, with the main 
improvements being a larger dataset (110~pb$^{-1}$) and a better determination
of the systematic uncertainties. This measurement used a sample of events
with a lepton, $\Etmiss$, at least three jets with $E_T > 15$~GeV and 
$|\eta|<2$, and a fourth jet
with $E_T > 8$~GeV and $|\eta|<2.4$. Events were further required to 
contain an SVX- or SLT-tagged jet. Thirty-four such events had an acceptable
$\chi^2$ when fit to the $\ttbar$ hypothesis, with a calculated background
of $6.4^{+2.1}_{-1.4}$ events.

This technique, while powerful, does not take account of all the available
information. It does not exploit the difference in signal to background 
between SVX tags and SLT tags, nor does it use any information from 
untagged events that satisfy the kinematic requirements for top.  
CDF has recently completed an optimized mass analysis that takes full 
advantage of this information.

To determine the optimal technique for measuring the mass, 
Monte Carlo samples of signal and background events are generated and
the selection cuts for the mass analysis are applied. This sample is then 
divided into several 
nonoverlapping subsamples, in order of decreasing signal to background:
SVX double tags, SVX single tags, SLT tags (no SVX tag), and untagged
events.
The mass resolution for each subsample is obtained by performing many 
Monte Carlo ``pseudo-experiments.''
Each pseudo-experiment for a given subsample contains the number of events 
observed in the data, with the number of background events thrown according 
its predicted mean value and uncertainty. For example, 15 SVX single-tagged
events are observed in the data, so the pseudo-experiments for the ``single
SVX-tag'' channel each contain 15 events, with the number of background
events determined by Poisson-fluctuating the estimated background in this
channel of $1.5\pm 0.6$ events. The standard likelihood fit to
top plus background templates is then performed for each pseudo-experiment.
The mass resolutions are slightly different for each subsample 
because single-, double-, and untagged events have different combinatorics,
tagger biases, etc. Top mass templates are therefore generated for each
subsample. By performing many pseudo-experiments, CDF obtains the expected
statistical error for each subsample.

Because the subsamples are nonoverlapping by construction, the likelihood
functions for each subsample can be multiplied together to yield a combined
likelihood. Monte Carlo studies have been performed to determine which
combination of subsamples produces the smallest statistical error. One
might expect that the samples with SVX tags alone would yield the best
measurement, because of their high signal to background. However it turns
out that the number of events lost by imposing this tight tagging requirement
more than compensates for the lower background, and actually gives a 
slightly larger statistical uncertainty than the previous CDF technique
of using SVX or SLT tags. Instead, the optimization studies show that
the best measurement is obtained by combining double SVX tags, single SVX tags,
SLT tags, and untagged events.
For the untagged events, these Monte Carlo studies show that a smaller
statistical error results from requiring the fourth jet to satisfy the
same cuts as the first three jets, namely $E_T>15$~GeV and $|\eta|<2$. For
the various tagged samples, the fourth jet can satisfy the looser requirements
$E_T>8$~GeV, $|\eta|<2.4$. The median statistical error expected from
combining these four samples is 5.4~GeV, compared to 6.4~GeV expected from the
previously-used method. This reduction in statistical uncertainty is
equivalent to increasing the size of the current SVX or SLT tagged data
sample by approximately 40\%.

\begin{figure}[htbp]
\begin{center}
\leavevmode
\epsfysize=3.5in
\epsffile{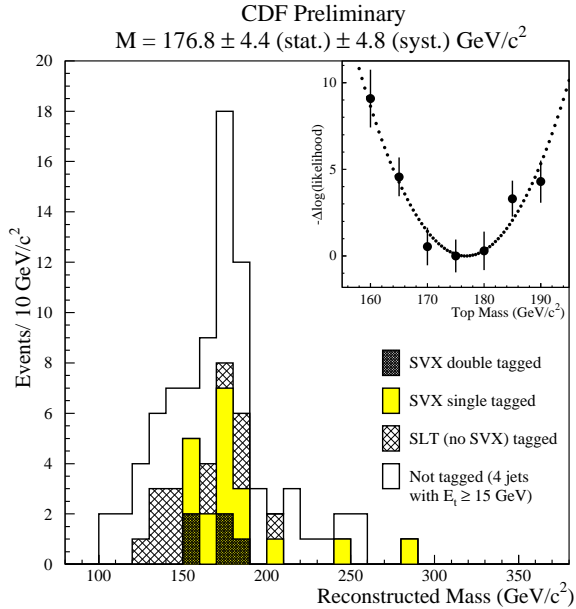}
\caption{Top mass distribution for all four of the CDF subsamples combined.}
\label{fig:mt_opt_like}
\end{center}
\end{figure}

The optimized procedure is then applied to the lepton plus jets
data. Table~\ref{tab:mass_subsamples} shows the number of observed events
in each subsample, together with the expected number of signal and
background events, the fitted mass, and the statistical uncertainties.
The result is $M_{top} = 176.8\pm 4.4$(stat)~GeV. The statistical uncertainty
is somewhat better than the 5.4~GeV expected from the pseudo-experiments. 
Approximately 8\% of the pseudo-experiments have a statistical uncertainty of 
4.4~GeV or less, so the data are within expectations. 
Figure~\ref{fig:mt_opt_like} shows the reconstructed mass distribution
for the various subsamples, together with the results of the fit.

\begin{table}[htbp]
\begin{center}
\caption{Mass-fit subsamples for the CDF top mass measurement. The first
row gives the results from the method of Refs.~\protect\cite{cdf_obs}
and \protect\cite{top_evidence_prd}. The next four rows show
the results from the subsamples used in the optimized method. The last
row shows the results of combining the four subsamples.}
\begin{tabular}{lccc}
\hline\hline
Subsample       & $N_{obs}$ & $N_{bkgd}$ & Fit Mass  \\
                &           &            &  (GeV)     \\ \hline
SVX or SLT tag  & 34  & $6.4^{+2.1}_{-1.4}$ & $175.6 \pm 5.7$   \\
(Prev. Method) &         &            &            \\ \hline
SVX double tag  & 5   & $0.14\pm 0.04$ & $174.3  \pm 7.9$   \\ 
SVX single tag  & 15  & $1.5\pm 0.6$   & $176.3  \pm  8.2$   \\ 
SLT tag (no SVX)& 14  & $4.8\pm 1.5$   & $140.0  \pm  24.1$  \\ 
Untagged ($E_T^4>15$)& 48 & $29.3\pm 3.2$ & $180.9 \pm 6.4$   \\ \hline
Optimized Method &    &         & $176.8 \pm 4.4$  \\ \hline

\end{tabular}
\label{tab:mass_subsamples}
\end{center}
\end{table}

Systematic uncertainties in the CDF measurement are summarized in 
Table~\ref{tab:cdf_mt_systs}. The largest systematic is the combined
uncertainty in the jet $E_T$ scale and the effects of soft gluons ({\em i.e.}
fragmentation effects). Such effects include calorimeter nonlinearities
and cracks, the effect of the underlying event, and Monte Carlo modeling
of the jet energy flow outside the clustering cone. The ``hard gluon''
systematic comes from the uncertainty in the fraction of $\ttbar$ events 
where one of the four highest-$E_T$ jets is a gluon jet from initial- or
final-state radiation. The \textsc{Herwig} Monte Carlo program predicts that 
55\% of the time a gluon jet is among the four leading jets.
This systematic is evaluated by varying the fraction
of such events by $\pm 30\%$ in the Monte Carlo 
and determining the resulting
mass shift. Systematics from the kinematic and likelihood fit are determined 
by using slightly different but equally reasonable methods of performing
the constrained fit and the final likelihood fit for the top mass. Such 
variations include 
allowing the background to float, or varying the range over which the
parabolic fit that determines the minimum and width of the likelihood 
function is performed. The ``different MC generators'' systematic is assigned 
by generating the $\ttbar$ templates with {\sc Isajet} instead of the default
{\sc Herwig}. Systematics in the background shape are evaluated by varying the
$Q^2$ scale in the Vecbos Monte Carlo program that models the
$W$ + jets background. Studies have shown that the relatively small non-$W$
background is kinematically similar to $W$ + jets. The systematic from
$b$-tagging bias includes uncertainties in the jet $E_T$-dependence of 
the $b$-tag efficiency and fake rate, and in the rate of tagging non-$b$
jets in top events. Monte Carlo statistics account for the remainder of
the systematic uncertainties. The final result is:
\begin{equation}
      M_{top}= 176.8\pm 4.4 ({\rm stat})\pm 4.8 ({\rm syst})~{\rm GeV}~({\rm 
                                                          CDF~prelim.}) 
\end{equation}

\begin{table}[htbp]
\begin{center}
\caption{Systematic uncertainties in the CDF top mass measurement.}
\begin{tabular}{cc}
\hline\hline
Systematic                   &       Uncertainty (GeV) \\ \hline

Soft gluon + Jet $E_T$ scale &          3.6 \\
Hard gluon effects           &          2.2 \\
Kinematic \& likelihood fit  &          1.5 \\
Different MC generators      &          1.4 \\
Monte Carlo statistics       &          0.8 \\
Background shape             &          0.7 \\
$b$-tagging bias             &          0.4 \\ \hline
Total                        &          4.8 \\ \hline
\end{tabular}
\label{tab:cdf_mt_systs}
\end{center}
\end{table}

\section{Kinematic Properties}
\label{sec-kine}

\begin{figure}[htbp]
\begin{center}
\leavevmode
\epsfysize=3.5in
\epsffile{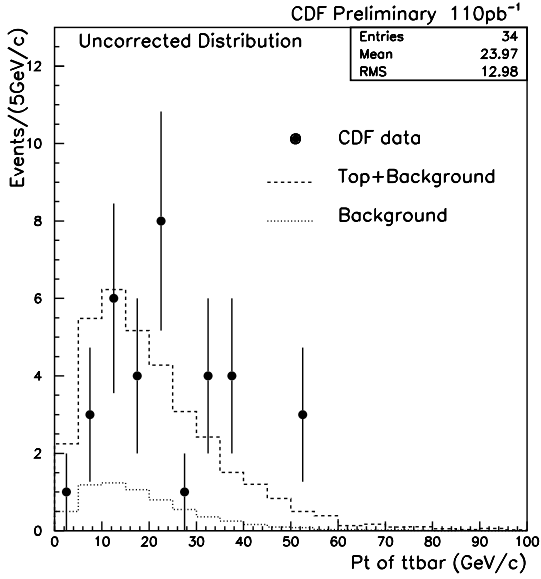}
\caption{Reconstructed $P_T$ of the $\ttbar$ system.}
\label{fig:kin_Pt_ttbar}
\end{center}
\end{figure}

\begin{figure}[htbp]
\begin{center}
\leavevmode
\epsfysize=3.5in
\epsffile{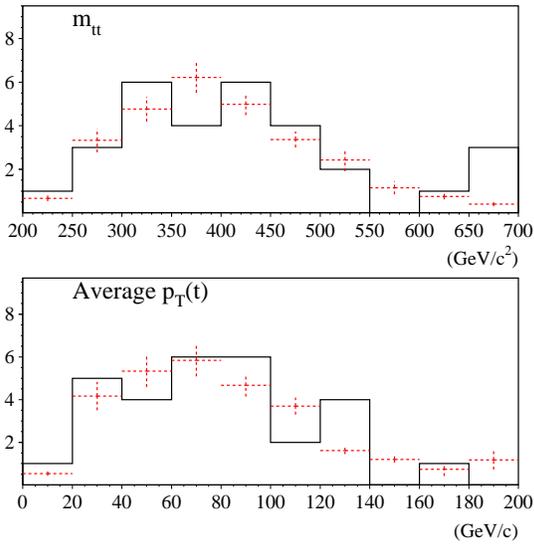}
\caption{Reconstructed $\ttbar$ invariant mass (top) and average $t$ or
$\bar{t}$~$P_T$ (bottom) from D0 data.}
\label{fig:kin_D0_Mtt_Pttop}
\end{center}
\end{figure}

The constrained fits described above return the complete four-vectors 
for all the partons in the event, and allow a range of other kinematic
variables to be studied. As examples, Fig.~\ref{fig:kin_Pt_ttbar}
shows the $P_T$ of the $\ttbar$ system as reconstructed from CDF data,
and Fig.~\ref{fig:kin_D0_Mtt_Pttop} shows the $\ttbar$ invariant mass
and the average $t$ and $\bar{t}$ $P_T$ from D0. The distributions have
not been corrected for event selection biases or combinatoric misassignments.
In these and in similar plots, the agreement with the Standard Model
is good.

\begin{figure}[htbp]
\begin{center}
\leavevmode
\epsfysize=3.5in
\epsffile{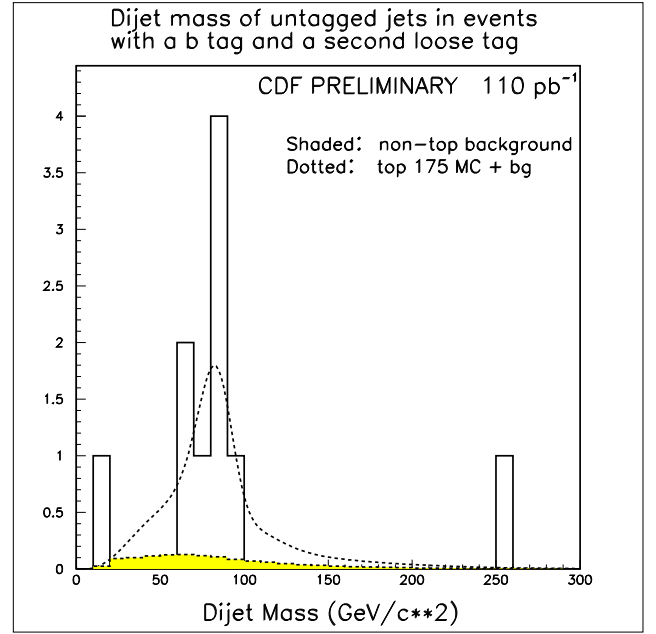}
\caption{Reconstructed hadronic $W$ peak in double-tagged top 
         candidate events.}
\label{fig:kin_mjj_dbl_tag}
\end{center}
\end{figure}

A very important cross-check that the experiments are really observing
$\ttbar$ pair production is to search for the hadronically decaying
$W$ in lepton + jets events. CDF has performed such an analysis by selecting
lepton + 4-jet events with two $b$-tags. To maximize the $b$-tag efficiency,
the second $b$ in the event is allowed to satisfy a looser tag requirement.
The two untagged jets should then correspond to the hadronic $W$ decay.
Fig.~\ref{fig:kin_mjj_dbl_tag} shows the dijet invariant mass for the
two untagged jets. The clear peak at the $W$ mass, together with the 
lepton, the $\Etmiss$, and the two tagged jets, provides additional
compelling evidence that we are observing $\ttbar$ decay to two $W$'s
and two $b$'s. 
This measurement is also interesting because it suggests
that in future high-statistics experiments the jet energy scale can be 
determined directly from the data by reconstructing this resonance.

\section{Branching ratios, ${V_{\lowercase{tb}}}$}
\label{sec-Vtb}

In the Standard Model, the top quark decays essentially 100\% of the time
to $Wb$. Therefore the ratio of branching ratios
\begin{equation}
          B = \frac{BR(t\rightarrow Wb)}{BR(t\rightarrow Wq)},
\end{equation}
where $q$ is any quark, is predicted to be one. CDF has measured $B$
using two techniques. The first technique compares the ratio of
double- to single-tagged lepton + jets events that pass the mass analysis 
cuts, and double-, single- and
un-tagged dilepton events. Since the efficiency to tag a single $b$-jet
is well known from control samples, the observed tag ratios can be
converted into a measurement of $B$. CDF finds:
\begin{equation}
    B = 0.94 \pm 0.27~(\mathrm{stat})~\pm 0.13~(\mathrm{syst}),
\end{equation}
or
\begin{equation}
    B > 0.34 ~\mathrm{(95\%~C.L.)}
\end{equation}
Untagged lepton + jets events are not used in this analysis because of the
large backgrounds admitted by the standard cuts. (Of course, the cuts were
designed to be loose to avoid kinematic bias; the background rejection is
normally provided by $b$-tagging.) The second CDF technique uses the
``event structure'' cuts of Ref.~\cite{top_kin_prd} to increase the purity
of the untagged lepton + jets sample, allowing it to be included in this 
measurement. The result is:
\begin{equation}
    B = 1.23^{+0.37}_{-0.31},
\end{equation}
or
\begin{equation}
    B > 0.61 ~\mathrm{(95\%~C.L.)}
\end{equation}
It should be noted that these analyses make the implicit assumption that the
branching ratio to non-$W$ final states is negligible. The fact
that the cross sections measured in the dilepton, lepton + jets, and 
all-hadronic channels are in good agreement is evidence that this assumption
is correct. Alternatively, if one believes the theoretical cross section,
it is clear from the SVX and SLT $b$-tag measurements that this cross section 
is saturated by decays to $Wb$. However, these ``indications'' have not
yet been turned into firm limits on non-$W$ decays.

The measurement of $B$ above can be interpreted as a measurement of
the CKM matrix element $V_{tb}$. However, it is not necessarily the case
that $B=1$ implies $V_{tb}=1$. This inference follows only in the absence of a 
fourth generation, where the value of $V_{tb}$ is constrained by unitarity 
and the known values of the other CKM matrix elements. In this case, 
$V_{tb}$ is determined much more accurately from these constraints
than from the direct measurement. (In fact, under the assumption of 
3-generation unitarity, $V_{tb}$ is actually the \textit{best known}
CKM matrix element.) A more general relationship, which is true for
three or more generations provided that there is no fourth generation
$b^{\prime}$ quark lighter than top, is
\begin{equation}
 B=\frac{BR(t\rightarrow Wb)}{BR(t\rightarrow Wq)} = 
   \frac{|V_{tb}|^2}{|V_{td}|^2 + |V_{ts}|^2 + |V_{tb}|^2}.
\label{eqn:Vtb}
\end{equation}

Since $B$ depends on three CKM matrix elements and not just one, a single
measurement cannot determine $V_{tb}$, and we must make additional assumptions
about $V_{ts}$ and $V_{td}$. 
In general, a fourth generation would allow $V_{td}$ and $V_{ts}$ to take on
any value up to their values assuming 3-generation unitarity.
One simplifying assumption is that the upper 3$\times 3$ portion of the CKM 
martix is unitary.
In that case, $|V_{td}|^2 + |V_{ts}|^2 + |V_{tb}|^2 = 1$, and
$B$ gives $V_{tb}$ directly. However, as noted above, under this assumption
$V_{tb}$ is very well determined anyway and this direct measurement adds no
improved information. Assuming 3$\times 3$ unitarity, the two analyses
described above give $V_{tb} = 0.97\pm0.15~\mathrm{(stat)}\pm
0.07~\mathrm{(syst)}$ and $V_{tb} = 1.12\pm0.16$ respectively. A more 
interesting assumption is that 3$\times 3$ unitarity is relaxed 
\textit{only} for $V_{tb}$. Then we can insert the PDG values of $V_{ts}$
and $V_{td}$ and obtain:
\begin{equation}
   V_{tb} > 0.022~\mathrm{(95\%~C.L.)}
\end{equation}
for the first method, or
\begin{equation}
   V_{tb} > 0.050~\mathrm{(95\%~C.L.)}
\end{equation}
for the second.

To see that a small value of $V_{tb}$ would not violate anything we know
about top, consider the situation with $b$ decays. The $b$ quark decays
$\approx 100\%$ of the time to $Wc$, even though $V_{cb}\approx 0.04$.
This is because the channel with a large CKM coupling, $Wt$, is
kinematically inaccessible. The same situation could occur for top in
the presence of a heavy fourth generation. However in this case the top
width would be narrower than the Standard Model expectation.
A more definitive measurement\cite{willenbrock}
of $V_{tb}$ will be performed in future Tevatron runs by measuring
$\Gamma_{t\rightarrow Wb}$ directly through the single top production 
channel $p\bar{p}\rightarrow W^{*} \rightarrow t\bar{b}$.

\section{Rare Decays}
\label{sec-rare}

CDF has performed searches for the flavor-changing neutral current decays
$t\rightarrow qZ$ and $t\rightarrow q\gamma$. The decay to $qZ$ can have
a branching ratio as high as $\sim 0.1\%$ in some theoretical 
models\cite{fritzsch}.
The search for this decay includes the possibility that one or both
top quarks in an event can decay to $qZ$. In either case the signature
is one $Z\rightarrow ll$ candidate and four jets.
Backgrounds in the $qZ$ channel come from $WZ$ and $ZZ$ plus jets
production, and are estimated to be $0.60\pm 0.14 \pm 0.12$ events.
In addition, 0.5 events are expected from Standard Model $\ttbar$ decay.
One event is observed.
Under the conservative assumption that this event is signal, the resulting
limit is:
\begin{equation}
   BR(t\rightarrow qZ) < 0.41~\mathrm{(90\%~C.L.)}
\end{equation}

The branching ratio of $t\rightarrow q\gamma$ is predicted to be roughly
$10^{-10}$\cite{parke}, so any observation of this decay would probably
indicate new physics. CDF searches for final states in which one top
decays to $Wb$ and the other decays to $q\gamma$. The signature is
then $l\nu\gamma + 2$ or more jets (if $W\rightarrow l\nu$), or 
$\gamma + 4$ or more jets (if $W\rightarrow jj$). In the hadronic channel,
the background is 0.5 events, and no events are seen. In the leptonic
channel, the background is 0.06 events, and one event is seen. (It is 
a curious event, containing a 72~GeV muon, an 88~GeV $\gamma$ candidate,
24~GeV of $\Etmiss$, and three jets.) Conservatively assuming this event
to be signal for purposes of establishing a limit, CDF finds:
\begin{equation}
   BR(t\rightarrow q\gamma) < 0.029~\mathrm{(95\%~C.L.)}
\end{equation}
This limit is stronger than the $qZ$ limit because of the $Z$ branching
fraction to $ee + \mu\mu$ of about 6.7\%, compared to the $\gamma$ 
reconstruction efficiency of about 80\%.

\section{W Polarization}
\label{sec-Wpol}

The large mass of the top quark implies that the top quark decays before
hadronization, so its decay products preserve the helicity structure
of the underlying Lagrangian. Top decays, therefore, are a unique
laboratory for studying the weak interactions of a bare quark. In particular,
the Standard Model predicts that top can only decay into left-handed or
longitudinal $W$ bosons, and the ratio is fixed by the relationship
\begin{equation}
\frac{W_{long}}{W_{left}} = \frac{1}{2}\frac{M_t^2}{M_W^2}.
\end{equation}
For $M_t = 175$~GeV, the Standard Model predicts that about 70\% of top
quarks decay into longitudinal $W$ bosons. 
This is an exact prediction resulting from Lorentz invariance and the $V-A$
character of the electroweak Lagrangian. If new physics modifies the
$t$-$W$-$b$ vertex---i.e. through the introduction of a right-handed 
scale---it may reveal itself in departures of the $W$ polarization from 
the Standard Model prediction. The $W$ polarization has recently been
measured, albeit with low statistics, by CDF. I describe this measurement
here to illustrate the type of measurement that will be done with high
precision in future runs with the Main Injector.

The $W$ polarization is determined from the $\cos\theta^{*}_l$, the
angle between the charged lepton and the $W$ in the rest frame of the $W$.
This quantity can be expressed in the lab frame using the approximate
relationship\cite{kane_wpol}
\begin{equation}
  \cos\theta^{*}_l \approx \frac{2m_{lb}^2}{m_{l\nu b}^2 - M_W^2},
\end{equation}
where $m_{lb}$ is the invariant mass of the charged lepton and the
$b$ jet from the same top decay, and $m_{l\nu b}$ is the three-body invariant
mass of the charged lepton, the neutrino, and the corresponding $b$ jet.
This last quantity is nominally equal to $M_t$, though in the analysis the
measured jet and lepton energies are used, and the possibility of combinatoric 
misassignment is included. 

Monte Carlo templates for $\cos\theta^{*}_l$ are generated using the 
{\sc Herwig}
$\ttbar$ event generator followed by a simulation of the CDF detector. The 
simulated events are then passed through the same constrained fitting
procedure used in the top mass analysis. The fit is used here to determine
the most likely jet--parton assignment (i.e. which of the two $b$ jets to
assign to the leptonic $W$ decay), and to adjust the measured jet and 
lepton energies within their uncertainties in order to 
obtain the best resolution on $\cos\theta^{*}_l$. The same procedure
is applied to $W$+jets events generated by the Vecbos Monte Carlo
program to obtain the $\cos\theta^{*}_l$ distribution of the background.

\begin{figure}[htbp]
\begin{center}
\leavevmode
\epsfysize=3.5in
\epsffile{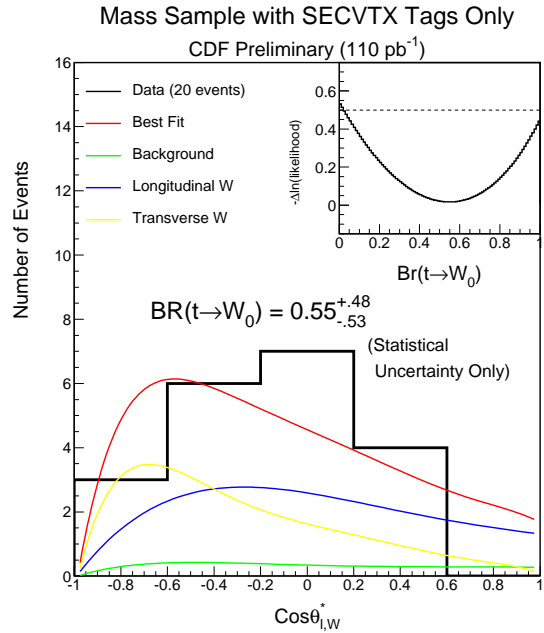}
\caption{Results of fit to the $\cos\theta^{*}_l$ distribution, used to determine the $W$ polarization in top decays. The dataset is the CDF top mass sample with only SVX tags allowed.}
\label{fig:wpol}
\end{center}
\end{figure}

The $\cos\theta^{*}_l$ distribution from the data is then fit to a 
superposition of Monte Carlo templates to determine the fraction of
longitudinal $W$ decays. The dataset is the same as in the CDF top mass
analysis (lepton + $\Etmiss$ + three or more jets with $E_T>15$~GeV and
$|\eta|<2$, and a fourth jet with $E_T>8$ GeV and $|\eta|<2.4$). To increase
the purity, only events with SVX tags are used. The $\cos\theta^{*}_l$
distribution in this sample is shown in Fig.~\ref{fig:wpol} together with 
the results of the fit. The fit returns a longitudinal $W$ fraction of 
$0.55^{+0.48}_{-0.53}$ (statistical uncertainties only). The statistics are
clearly too poor at present to permit any conclusions about the structure of
the $t$-$W$-$b$ vertex. However, with the large increase in statistics
that the Main Injector and various planned detector improvements will 
provide, precision measurements of this vertex will become possible.
Studies indicate, for example, that with a 10~fb$^{-1}$ sample one can measure 
BR($t\rightarrow W_{long}$) with a statistical uncertainty of about 
2\%, and have sensitivity to decays to right-handed $W$'s with a 
statistical precision of about 1\%\cite{TEV2000}.

\section{Conclusions}
\label{sec-concl}

The Tevatron experiments have progressed quickly from the top search 
to a comprehensive program of top physics. Highlights of  the recently
completed run include measurements of the top cross section and mass,
studies of kinematic features of top production, and a first look at the
properties of top decays. Many of these analyses are still in progress,
and improved results can be expected. 

With a mass of approximately 175~GeV, the top quark is a unique
object, the only  known fermion with a mass at the natural electroweak
scale. It would be surprising if the top quark did not play a role in
understanding electroweak symmetry breaking. Current measurements are all
consistent with the Standard Model but in many cases are limited by
poor statistics: the world $\ttbar$ sample numbers only about a hundred
events at present. Both CDF and D0 are undertaking major detector upgrades 
designed to take full advantage of high-luminosity running with the Main 
Injector starting in 1999. This should increase the top sample by a 
factor of $\sim$50. Beyond that, Fermilab is considering plans to 
increase the luminosity still further, the LHC is on the horizon, and
an $e^+e^-$ linear collider could perform precision studies at the $\ttbar$
threshold. The first decade of top physics has begun, and the future
looks bright.

\section{Acknowledgements}

I would like to thank the members of the CDF and D0 collaborations whose
efforts produced the results described here, and the members of the
technical and support staff for making this work possible. 
I would also like to thank the organizers of Snowmass `96 for inviting
me to present these results and for sponsoring a stimulating and enjoyable
workshop. This work is supported in part by NSF grant PHY-9515527 and by a DOE
Outstanding Junior Investigator award.

%%%%% References
%

\end{document}